\documentclass[prb,aps,twocolumn,showpacs,superscriptaddress]{revtex4-1}
\usepackage{hyperref}
\hypersetup{colorlinks, citecolor=blue, filecolor= black, linkcolor= black, urlcolor= blue}
\usepackage{graphicx}
\graphicspath{ {phase-diag/} {regimes/} }
\usepackage{dcolumn}
\usepackage{bm}
\usepackage{float}
\usepackage{bm}
\usepackage{epsfig}
\usepackage{latexsym}
\usepackage{amsmath}
\usepackage{mathtools}
\usepackage{amsfonts}
\usepackage{color}
\usepackage{array}
\usepackage{enumitem}
\usepackage{hyperref}
\hypersetup{
    colorlinks=true,
    linkcolor=blue,
    filecolor=magenta,      
    urlcolor=cyan,
}
\usepackage{lipsum}
\usepackage{amssymb, amsmath}

\DeclarePairedDelimiterX\braket[2]{\langle}{\rangle}{#1 \delimsize\vert #2}

\newcommand{\<}{\langle}

\renewcommand{\>}{\rangle}
\renewcommand{\(}{\left(}
\renewcommand{\)}{\right)}

\renewcommand{\d}{\partial}

\begin{document}
	\preprint{Preparing}
	
	\title{Partial condensation of mobile excitons in graphene multilayers}
	
	\author{Igor V. Blinov}
	\email{blinov@utexas.edu}
	\affiliation{
		Department of Physics, The University of Texas at Austin, Austin, Texas 78712, USA}
		\author{Chunli Huang}
	\affiliation{
		Department of Physics, The University of Texas at Austin, Austin, Texas 78712, USA}
		
	\author{Nemin Wei}
	\affiliation{
		Department of Physics, The University of Texas at Austin, Austin, Texas 78712, USA}

	\author{Qin Wei}
	\affiliation{
		Department of Physics, The University of Texas at Austin, Austin, Texas 78712, USA}

				\author{Tobias Wolf}
	\affiliation{
		Department of Physics, The University of Texas at Austin, Austin, Texas 78712, USA}

				\author{Allan H. MacDonald}
	\affiliation{
		Department of Physics, The University of Texas at Austin, Austin, Texas 78712, USA}

	\date{July 3, 2022}
	
	\begin{abstract}
		\noindent
		At a large displacement field, in rhomboedral and Bernal-stacked graphene a normal paramagnetic state transitions to a correlated state. Recent experiments showed that such systems have several phase transitions as a function of the carrier density. The phase adjacent to a paramagnetic state has anomalously high resistance and reduced degeneracy of the Fermi sea. We show that both phenomena can be explained through a concept of partial intervalley exciton condensation: a fraction of particles condenses into excitons, and another forms an intervalley coherent Fermi liquid. The exciton part of the system do not contribute to the electrical current thus increasing the resistance. Within this paradigm, the increase in the resistance has entirely geometrical origin. We check validity of the phenomenological theory through numerical calculations. We also show that the quantum oscillation data should not be very different between the partial excitonic state and the intervalley coherent states suggested by other authors. Further, we suggest STM/AFM or Raman spectroscopy to have a conclusive evidence for the occurrence of the partial exciton condensation that we suggest in this paper. 
		\end{abstract}

	\maketitle
	\newpage
	
	
\textit{Introduction}
A long held belief was that two-dimensional systems are intrinsically unstable \cite{Landau1937phase,mermin1968crystalline} because thermal fluctuations will unavoidably destroy any kind of long-range order. However, further research of the two-dimensional systems beyond harmonic approximation showed that in principle, slight bending in the third dimension could stabilize structures. Non-negligible anharmonicity present in the graphene because of the strong carbon bonds couples the deformation in the third dimension to in-plane long-range phonons\cite{meyer2007structure, nelson1987fluctuations}. As a result, quasi-two-dimensional sheet can, at least theoretically, become stable \cite{zakharchenko2009finite}. Later, it was demonstrated experimentally \cite{neto2009electronic,novoselov2004electric,zhu2010graphene} that stable (or quasi-stable) atomically-thick structures of graphene can be obtained from graphite through exfoliation \cite{novoselov2005two}. While full of defects, both encapsulated and free-standing graphene showed sufficient stability under a wide range of external conditions.
 Originating electron quasiparticles in graphene while have an exotic quasi-relativistic dispersion relation \cite{slonczewski1958band,wallace1947band}, do not show interaction-induced collective phases \cite{kotov2012electron} for low densities in the absence of the magnetic field. 

When several layers of graphene put in contact, hybridization between them flattens the band structure, effectively generating mass for the low-energy quasiparticles. As a result, because of the increase in the density of states electrons in graphene multilayers, correlated phases in multilayer do occur \cite{kotov2012electron,weitz2010broken}. 

Further, a low-energy bands can be flattened even more, if two graphene sheets are rotated by a small angle $\theta$ with respect to each other \cite{bistritzer2011moire}.  

So called moir\'e materials drawn a lot of attention because of the rich variety of the symmetry broken phases \cite{cao2018unconventional,yankowitz2019tuning,arora2020superconductivity} present in the system and even superconductivity of yet unknown origin \cite{wu2018theory, lian2019twisted, kozii2019nematic, chichinadze2020nematic, khalaf2020charged}. Such materials are hard to make and suffer of large disorder inevitably present in the system because of preparation process. Another problem lies in the theoretical realm: large number of bands $\propto1/\theta \approx 10^2$ present in the system makes mathematical description complicated and, as a result, there are still gaps in understanding.

A new hope was brought by recent experiments in ABC-stacked graphene trilayers \cite{zhou2021half,zhou2021superconductivity} and AB-bilayers \cite{seiler2022quantum}. Canonical description of the non-interacting model at low doping consists of 6 and 4 bands correspondingly for each spin and valley making theoretical description much easier than in the moir\'e systems.  Both systems, trilayer and bilayer, at a non-zero perpendicular electrical field show a cascade of phase transitions as a function of hole/electron density and displacement field. Far away from the charge neutrality the system is fully symmetric in $SU(2)\times SU(2)$ isospin space (paramagnetic phase) and, as a result, Fermi sea is 4-fold degenerate. 

The first transition on the hole-doped side happens at large ($>0.2$V/nm) non-zero displacement field and a hole density of order $0.3\times 10^{12}$ $\text{cm}^{-2}\approx 2\times 10^{-4} $ per unit cell per flavor. The transition was first seen in the quantum oscillation data\cite{zhou2021half}. As the system crosses from the paramagnetic phase to the first correlated phase, Fermi sea degeneracy is reduced by a factor of 2. This phase was dubbed as a partially isospin polarized (PIP) phase. At low temperatures, on the interface between PIP and the paramagnetic one, a superconducting phase arises \cite{zhou2021superconductivity}. While several mechanisms for the superconductivity are present \cite{you2022kohn,chatterjee2021inter,dong2021superconductivity}, its exact mechanism remains murky. One of the possible scenarios for the emergence of the superconducting phase is the appearance of the attractive interactions through the fluctuations \cite{dong2021superconductivity,chatterjee2021inter} of the order parameter of the first coherent phase. 

Coherent phase appears in a regime of hole doping with the two Fermi surfaces (annular Fermi surface) present within each flavor. Our belief is that this is the most important feature of the system and will exploit it in order to build a simple model and explain the observable features of the system. The symmetry broken phase does not show signatures of spin or valley polarization. As a candidate, an intervalley coherent phase was suggested \cite{chatterjee2021inter, huang2022spin, you2022kohn}.

However, such phase does not explain all observable facts. Namely, while coherence effectively reduces the number of Fermi surfaces, thus explaining the change in the quantum oscillations  such phase do not seem to explain \cite{priv-res} increase of the resistance \cite{zhou2021superconductivity}. 
We aim to build a theory that would explain both the change in the quantum oscillations and most importantly, enhancement of the system's resistance. 

\textit{Simple model}
Even though bilayer and trilayer have different band structures,
we focus on their similarity rather than a difference to put forward the idea that the origin of the first correlated present in both could be the same. The first transition reduces the degeneracy of the Fermi surface by a factor of 2, hence the minimal model should have two flavors. Second, two features we think are of the outmost importance are the annular Fermi surface and the trigonal warping. The simplest dispersion relation for two flavors (two valleys) that has both properties is 
\begin{equation}\label{model:dispersion}
	\hat{\epsilon}(p)=
	\( m p^2 /2+\lambda p^4/4\)\hat{\tau}_0+\Delta \hat{\tau}_3 p^3 \cos\(3\alpha_p\)/2,
\end{equation}
where $m>0$ and $\lambda<0$, $\tau_i$-s are the Pauli matrices in valley space. For negligibly small trigonal warping, Fermi momentums are  $p_{F\pm}=m/4\pm1/2\sqrt{(m/2)^2+\lambda\mu}$. The condition for the annular Fermi surface to exist then is $0<\mu<m^2/|4\lambda|$ and $\lambda<0$ (Fig. \ref{fig:band-structure}). In what follows, we use $m$ as an interaction scale, while $p$ is taken dimensionless $p=a_0 k$. However, fitting to the 6-band model shows that $m$ is approximately $5\times 10^3$ K. In what follows, we will use parameters more relevant for the trilayer than the bilayer.

Interactions, present in the system, can drive it to a symmetry broken phase.  We choose the momentum-independent $SU(2)$-symmetric interaction
\begin{equation}
	H=\sum_p\hat{\epsilon}_k\psi_k^\dagger \psi_k+\frac{\lambda}{S}
	\sum_{kk'}(\psi_k^\dagger \hat{\tau} \psi_k)\cdot(\psi_{k'}^\dagger \hat{\tau}\psi_{k'}),
\end{equation}
where $\psi_{k}=(c_{k+},c_{k-})$ are the two-dimensional spinors in the valley space, $c_{kv}$ are the annihilation operators at momentum $k$ and valley $v$, $\hat{\tau}=(\tau_x,\tau_y,\tau_z)$ is a vector of Pauli matrices. The opening of the second Fermi surface at $\mu=0$ gives a substantial increase in the density of states, and consequently, can help to drive the system across the transition by a Stoner-like scenario. 

When $\Delta=0$ the model is $SU(2)$ symmetric in valley space and, as a result, valley-polarized ($\<\psi_k^\dagger\tau_z\psi_k\>\neq0$) and valley-coherent ($\<\psi_k^\dagger\tau_{x,y}\psi_k\>\neq0$)  phases have the same energy. The non-zero trigonal warping, in turn, makes the valley coherence more preferable with energy difference scaling as $\propto \Delta^2$. 

\begin{figure}
\includegraphics[width=1\columnwidth]{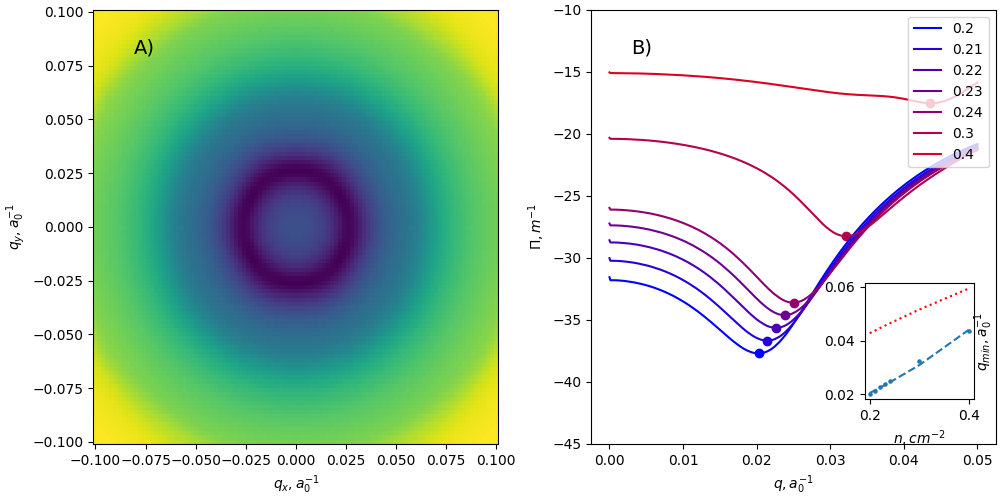}
\caption{A symmetrized response $\Pi(q)=\Pi_{+-}(q)+\Pi_{-+}(q)$ in the electron-hole channel between two different valleys. A) Two-dimensional colorplot for the response at the hole density $n_h=0.21\times 10^{12} cm^{-2}$. There is a noticeable minima around $q\approx p_{F+}-p_{F-}$. The real part of the response has a 6-fold symmetry. Unlike the response of the free electron gas it has a strong variation for the transferred momentum below $2 p_{F}$. B) Response $\tau(q)$ between the valleys for different values of the hole density.  The ratio of the $\Pi(q)/\Pi(0)$ reaches its  maxima at around $n_h=0.3 \times 10^{12} \ cm^{-2}$. On the inset: dependence of the position of the minima in the response on the density (blue dots), fit $p_{F+}-p_{F-}$ (blue dashed) and $(p_{F+}/v_{+}+p_{F-}/v_{-})/(v^{-1}_{+}-v^{-1}_{-})$.  In this calculation, parameters used are $m=1$, $\lambda=-240$, $\Delta=-0.8$. Fitting to the 6-band structure shows that $m$ of order $5\times 10^3$ K.   }
\label{fig:response}
\end{figure}

\textit{Intervalley response in the electron-hole channel } 
Usually, the homogenous solution of the mean-field equation corresponds to a more energetically favorable phase. However, in the small number of cases, usually when a discrepancy between the two Fermi surfaces is present, a phase with spontaneously broken spatial symmetry can arise. The likelihood of such phase is indicated by the peak in the undressed response in the channel of interest at a non-zero transferred momentum.  For a system with dispersion \eqref{model:dispersion}, whenever the two Fermi surfaces are present, the response in the intervalley electron-hole channel $\tau(\bar{q})$ has a minimum (Fig. \ref{fig:response}) at a momentum approximately equal to the difference between the Fermi momentums averaged over the angle.
\begin{figure}
\includegraphics[width=1\columnwidth]{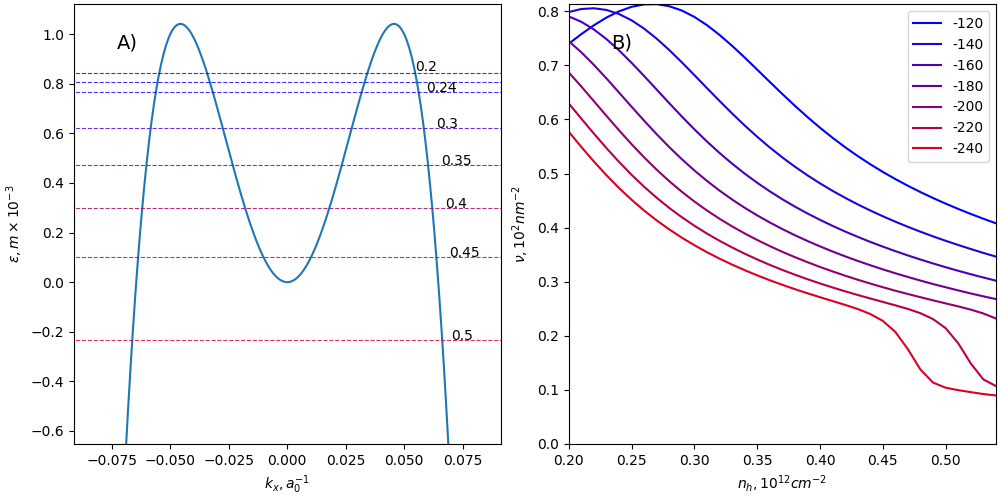}
\caption{Simple model \eqref{model:dispersion} that could have an annular Fermi surface and trigonal warping. A) Band dispersion for parameters $m=1$, $\Delta=-0.8$, $\lambda=-240$ in units of mass with Fermi energies for densities in the range $0.2$ to $0.5\times 10^{12}$ $cm^{-2}$ per valley per spin. In the experiment, first correlated phase shows for when the hole doping decreases down to $0.3\times 10^{12}$ $cm^{-2}$ for $0.23$ $V/nm$.
 B) Density of states as a function of hole density for $\lambda$ in the range $-120$ to $-240$. For $\lambda=-240$, there is a sharp jump at $n_h=0.45$ corresponding to the appearance of the second Fermi surface. }
\label{fig:band-structure}
\end{figure}

 In our case, the presence of the annulus means that the inner Fermi surface has an electron-like dispersion, and the outer Fermi surface has a hole-like dispersion. Then for an intervalley excitations in the electron-hole channel we could think of two different types of processes: excitations between the same Fermi surfaces, and excitations between two different Fermi surfaces. Response for excitations between the same Fermi surfaces is a monotonically increasing function of the transferred momentum. The smallest momentum at which the excitation between different Fermi surfaces at a vanishingly small transferred frequency become allowed is $\text{min}(p_{F+}(\theta_p)-p_{F-}(\theta_p))$. As the response between the Fermi surfaces is a monotonic function too, we should expect the enhancement of the response at the transferred momentum equal to the difference between two Fermi surfaces.

The minima in the undressed response means that the divergence in the RPA-response $(1+\lambda \tau^{-1}(q))^{-1}$ should arise earlier for non-zero transferred momentum, meaning that the symmetry broken state with particles in different valleys coupled through a non-zero momentum will potentially more stable than the quasi-homogenous state. The symmetry broken state will be characterized by the order parameter
\begin{equation}
	x_{vv'}(q)=\frac{1}{S}\sum_{p}\<c_{v'}(p)c^\dagger_{v}(p+q)\>,
\end{equation}
where $v\neq v'$ are the valley indices $+/-$.
If trigonal warping is present, the full rotational symmetry reduces to the $C_3$ symmetry: as a result, there is a set vectors $g_1,g_2,g_3$ with absolute value $q$ related by $C_3$-rotation where the instability is the strongest. We expect than the resulting mean-field state will be periodic with reciprocal lattice vectors equal to $g_1, g_2$ and $g_3$. 

\textit{Mean-field state} 
 We choose the reciprocal vectors to be $a_1=(q_{min},0)$, $C_3 a_1$, $C^2_3 a_1$, where $C_3$ is a rotation by $2\pi/3$. The resulting mean-field Hamiltonian has a form:
\begin{equation}\label{mean-field:hamiltonian}
	\hat{H}=\sum_{k} \psi^\dagger_{k}\hat{\epsilon}_{k} \psi_{k}+\sum_{k,g_{i}} \psi^\dagger_{k+g_{i}} \hat{x}_{g_i}  \psi_{k},
\end{equation}
where $\psi_{k}$ are the two-dimensional spinors, the coupling between two different valleys is $\hat{x}_q=\hat{\tau_x} (x_{+-}(q)+x_{-+}(q))/2+\hat{\tau}_y (x_{+-}(q)-x_{-+}(q))/2$.
For a Hamiltonian \eqref{mean-field:hamiltonian} a standard Bloch theorem is applicable. The initial Brillouin zone is then reduced to a significantly smaller (by a factor of $10^{-2}$) Brillouin zone. The corresponding period of the mean-field state solution $\propto 1/q$ is of order $10^2$ of the original lattice constants and depends on doping. The coupling at a finite momentum can give rise to oscillations in density that we discuss later in the paper. 

To find the value of the order parameter $x_q$ we use the weak-coupling approach expanding a mean-field equation up to the third order. Then
\begin{multline*}
    x_{+-}(g_i)=
           -\lambda x_{+-}(g_i)\Pi_{+-}(g_i)
           \\-
           \lambda \sum_{g_j,g_k}
        U_{+-}(g_i,g_j,-g_k)x_{+-}(g_j)x_{-+}(-g_k)x_{+-}(g_i-g_j+g_k),
\end{multline*}
where $\Pi_{+-}(g_i)$ is a simple polarization operator between '+' and '-' valleys at a momentum $g_i$, $\lambda$ is the interaction constant, $U_{+-}(g_i,g_j,-g_k)$ is a generalization of the polarization operator to the case of the four incoming/outgoing electron-hole excitations. The polarization operator $\Pi_{+-}(g_i)$ is invariant under $C_3$ rotation (but not under $C_6$ or the mirror reflection). Similarly,  $U_{+-}(g_i,g_j,-g_k)$ is invariant under simultaneous rotation of all vector over $C_3$ (but not under pairs or single vectors). As a result, order parameter should be invariant under $C_3$ (or, alternatively, some of the components vanish). Under mirror reflection  $x_{+-}(g_i)^*=x_{-+}(-g_i)$. Another restriction comes from the fact that the vector $g_i-g_j+g_k$ should have a length equal to $q$. Then at least two out of the three of the arguments in $U_{+-}$ should be the same. Because of the symmetry restrictions, three types of solutions are possible: 1) stripe solution with only a single component being non-zero, 2) symmetry broken with one of the three components vanishing, 3) $C_3$-symmetric solution. The mean-field equation reduces to
\begin{equation}
    |x_{+-}(g_i)|^2=
           -\frac{\lambda\Pi_{+-}(g_i)+1}{\lambda U_n(g_i)},
\end{equation}
where $\sum_{j,k} U_{+-}(g_i,g_j,g_k)\equiv U_n(g_i)$ is the sum over all terms that correspond to non-zero components of $x_q$. Then ratio of the energies for each solution with respect to the energy of the Fermi sea will depend only on the coefficients $U_n$ and not on the response:
\begin{equation}
 \frac{\Delta E_n}{\Delta E_k}=\frac{n |U_k|}{k |U_n|},
\end{equation}
where $n$ and $k$ is the number of the non-zero components. The direct process with $g_i=g_j=g_k$ have much higher amplitude and, as a result, the $\Delta E_n/\Delta E_k\approx n/k$. Then the symmetrical solution is more preferred then either 1) or 2). To distinguish the inter valley coherence at a finite momentum ($IVC-Q$) from other candidate phases and show that this is the phase that gives an explanation for the increase in the resistance and change in the quantum oscillation data, we now discuss the experimental features of the mean-field state.

\begin{figure}
\includegraphics[width=1\columnwidth]{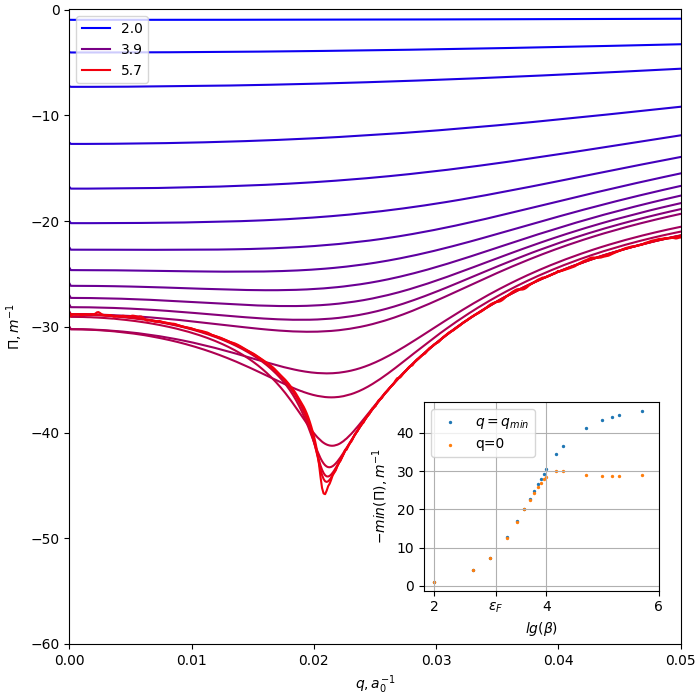}
\caption{Temperature dependence of the intervalley electron-hole response. Main plot: $\tau$ for several temperatures ranging from $\lg(\beta)=3.6$ to $\lg(\beta)=4$. One can see that at the minimum ($q=q_{min}$) the value of the response is by a factor of $1.5$ is larger than at $q=0$ by its absolute value. 
Inset: temperature-dependence of the minimum in the response in as a function of $\lg(\beta)$. 
 At the intermediate temperature response goes nearly logarithmically. Response at the $q=0$ at low ($10^{-4}<\epsilon_F$) temperatures. 
 In this calculation, parameters used are $m=1$, $\lambda=-240$, $\Delta=-0.8$, $n_h=0.21 \times 10^{12}  cm^{-2}$}. 
\label{fig:response:T-dep}
\end{figure}

\textit{Experimental features}
The first feature of the symmetry broken state to explain is an increase in the resistance.  A classical formula for the current in the leading order in the vector potential is $  j_i(i\Omega)=1`	
    \sum_\sigma \frac{\d H_{\sigma}}{\d A_i}$, where $\sigma=0,1$ is a spinor index and $i=x,y$ is a direction vector. As a result, $j_i(\Omega)=p_{ij}(\Omega)A_j(\Omega)=-ip_{ij}(\Omega)E_{j}(\Omega)/\Omega=\sigma(\Omega)_{ij} E_{j}$.
\begin{equation}\label{exp:response}
p_{ij}(\Omega)
=
\sum_{p n a b} v^i_{ab}(p) v^j_{ba}(p) G_a (\omega_n+\Omega,p)
G_b (\omega_n,p),
\end{equation}
where $a,b$ stand for bands, $i,j$ are the direction of the velocities matrix elements $v^i_{ab}\equiv \sum_{i\alpha}u_{a\alpha}(p+g_i)u^*_{b\alpha}(p+g_i)\d_i \epsilon(p+g_i)$, $G_{a/b}$ stands for the Green's funtions in the band basis averaged over the disorder.  Within the framework of the \eqref{exp:response}, it is the off-diagonal elements of the velocity that are responsible for the increase (Fig. \ref{fig:band-structure}, inset). Numerically, we have an increase of the resistance by the order of $10\%-30\%$ (Fig. \ref{fig:band-structure}) for $\beta=5\times10^4 m^{-1}$ and higher for lower temperatures. 

We explain the rise in the resistance in the following manner: the pairing momentum $q_{min}$ connects both points within the same Fermi surface as well as the points on different surfaces. The outer Fermi surface has a negative Fermi-velocity and thus hole-like, while the inner is electron-like. 

When the electrons and holes are paired, an exciton condensate can be formed. When electrons are paired in different valleys, a pseudo-magnetism (coherence) arises. As a result, we can think of the state as a mixture between a condensate of mobile excitons and a coherence of electrons in different valleys. The condensed part is, in principle, should have zero conductivity. For pure exciton condensate then $\tau=0$. Neglecting the change in the drift velocity, we expect the relative change in the total conductivity be proportional to the $-n_{ex}/n_{tot}$, where the $n_{ex}$ is the density of the particles in the condensate, $n_{tot}$ is the total electron density. The parameter $-n_{ex}/n_{tot}$ should be entirely geometrical. To estimate it, we do the following: the number of the exciton pairs should be proportional to the integral 
\begin{equation}
	n_x\propto \sum_i \int_{p_1} d^2 p_1 \int_{p_2}d^2 p_2 \delta(|\bar{p}_2-\bar{p}_1-\bar{q}_i|)
	n_i(p_2)n_o(p_1)/3,
\end{equation}
where the sum performed over the reciprocal lattice vectors, $n_i(p_1)=\theta(p_i-p_1)$, $n_o(p_2)=\theta(p_o-p_2)\theta(p_2-p_i)$, $\theta(x)$ is the Heaviside function, $p_{i/o}$ are the absolute values of the Fermi momentums of the inner and outer Fermi surfaces, $\delta(x)$ is a delta function. The integral over crescent-like area gives $n_x\propto 4\pi p_i^{3/2}q^{1/2}$ for small $q/p_i$. 
Similarly, the number of electrons in the system for a single spin flavor can be estimated as $n\propto 2\pi(p_o^2-p_i^2)$,
with $i/o$ corresponding to either inner or outer Fermi seas. As a result, conductance in the correlated regime divided by the conductance of the normal phase will give
\begin{equation}
	\sigma_x/\sigma_m\propto 1-\frac{2p_i^{3/2}q^{1/2}}{p_o^2-p_i^2}.
\end{equation}
The latter formula in the range of interest takes values from $0.2$ to $0.4$. 
\begin{figure}
\vspace{0.2 in}
\includegraphics[width=1\columnwidth]{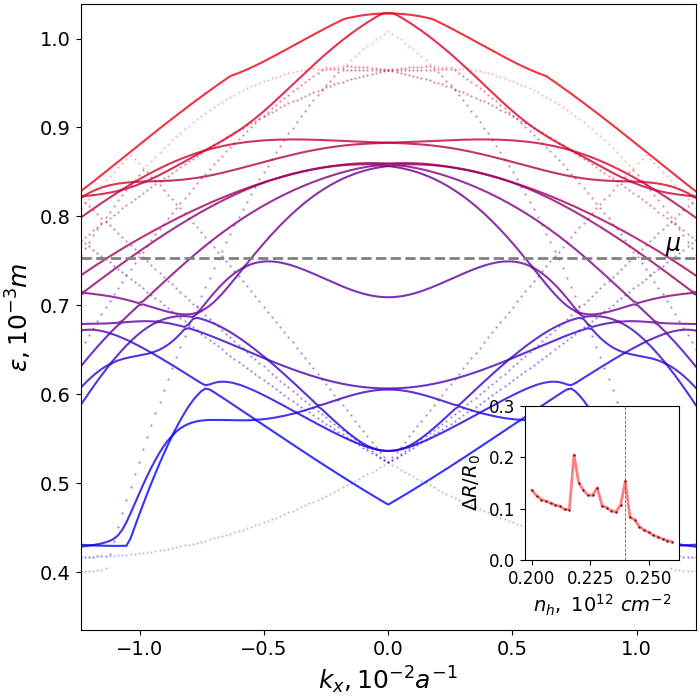}
\caption{Band structure of the mean-field solution at the density $n=0.24\times 10^{12}cm^{-2}$ close to the Fermi-level (solid lines) and the bands of the non-interacting model (dots). At this density, the value of the order parameter is $|x|=1.1\times 10^{-4} m$. Because the value of the order parameter $|x|$ is comparable to the bandwidth, the changes are drastic.  Inset: dependence of the relative increase of the resistance $\Delta R/R_0=(R_x-R_0)/R_x$, where $R_x$ is the resistance of the correlated phase, $R_0$ is the resistance of the paramagnetic phase. In this calculation, parameters used are $m=1$, $\lambda=-240$, $\Delta=-0.8$.  Temperature in this calculation is  $T=0.2 \times 10^{-4} m=10^{-1} K$. }. 
\label{fig:band-structure}
\end{figure}
To put this interpretation further, we study the temperature-dependence of the response (Fig. \ref{fig:response:T-dep}). Naively, the exciton part should scale as $\log(\beta)$, while the magnetism should saturate to a constant value at temperatures below $\epsilon_F$. We see the logarithmic behavior of the response for $q=q_{min}$ at the temperature up to two order of magnitude below the Fermi energy. At temperatures lower than that the response is constant. We could conclude then that at temperatures below the transition but not very low the response of the system effectively resembles one of the exciton condensate. However, as the temperature goes down, the fact that there is only a countable number of the points with perfect nesting becomes noticeable and the response of the system has a temperature behavior of the Stoner magnet. \\
The reduction of degeneracy of the Fermi surface seen in the quantum oscillation can be explained through the intervalley-coherence. In our case, the order parameter is inhomogeneous but has a periodicity of around $2 \ 10^2 a_0\approx  60$ nm. Given that the period is of the same order magnitude as the magnetic length or more, the corresponding matrix elements in Landau levels basis will only be non-zero between the same Landau levels. Thus, the IVC-Q will be approximately indistinguishable from the intervalley coherence at zero momentum for sufficiently large magnetic fields in the quantum oscillation data. 

To conclusively distinguish the phase with partial condensation of mobile excitons from the simple coherence between two valleys, we suggest looking at the density variations either through the STM, AFM or Raman spectroscopy. Unlike the Larkin-Ovchinnikov state, the variations in the order parameter do show up in the density. In the leading order in $x$, the amplitude of variations then can be estimated as $\delta n\approx 12 x^2\nu/\epsilon_q$. The latter makes approximately $1-10\% $ of the homogenous component. 

\textit{Summary}
We showed that a new state with partial condensation of mobile excitons can explain the observable increase in the resistance and quantum oscillation data. 
A theoretical possibility of such state is hinted by the peak in the electron-hole response between two valleys when the annular Fermi surfaces are present. Because of the extrema in the response at non-zero transferred momentum, the Fermi sea can become unstable towards transition to a state with intervalley coherence at finite momentum. We interpret the resulting phase of matter as a partial condensation of excitons distributed between two different Fermi surfaces (inner electron-like and outer hole-like) and coherence established between quasiparticles on the alike Fermi surface. The additional argument towards this interpretation is obtained through study of the temperature dependence of the response. We demonstrate that the bare response does not saturate even at temperatures significantly lower than the Fermi energy, with a divergence being slower than $\log(T)$. 
 
 Partial condensation also explains a significant increase in the resistance: part of the Fermi liquid that condenses into excitons and does not contribute to the electrical current. The fraction of the electrons that condenses is of order $2p_i^{3/2}q^{1/2}/(p_o^2-p_i^2)$ for $q/p_i\ll 1$ and $T\to 0$.  Further, we look at the long-range ($\approx 30 $ nm) density variations. We estimate the amplitude of the oscillations to be of order $1-10\%$ of the homogenous component and potentially observable through STM, AFM or Raman spectroscopy.  
 
\textit{Acknowledgement}
I.V.B. is grateful for helpful conversations with Anna M. Seiler and Noelia Fernandez, Haoxin Zhou as well as to a "lovely guy in a cowboy hat" who accepted this acknowledgement as an entrance fee to a swimming pool.
\bibliographystyle{apsrev4-1}
\bibliography{abc-bib.bib}

\end{document}